\documentclass[aps,prb,twocolumn,showpacs,showkeys,floatfix]{revtex4}

\usepackage{graphicx}
\graphicspath{{figs/}}
\begin{document}
\title{Bright and dark modes induced by graphene bubbles}
\author{Jin-Wu~Jiang}
    \altaffiliation{Electronic address: phyjj@nus.edu.sg}
    \affiliation{Department of Physics and Centre for Computational Science and Engineering,
             National University of Singapore, Singapore 117542, Republic of Singapore }

\author{Jian-Sheng~Wang}
    \affiliation{Department of Physics and Centre for Computational Science and Engineering,
                 National University of Singapore, Singapore 117542, Republic of Singapore }

\date{\today}
\begin{abstract}
Through a lattice dynamics analysis, it is revealed that the bubble plays a role of energy shield in the graphene, which helps to split the normal modes into two categories of distinct topological nature, namely the bright and dark modes. The topological invariants, Euler characteristic, of the bright and dark modes are 1 and 0, respectively. For bright modes, the energy is confined inside the bubble, so this type of modes are sensitive to the shape of the bubble; while opposite phenomenon is observed for the dark modes. The different behavior from these two types of normal modes is examined and verified in the process of phonon thermal transport. The bright and dark modes are expected to be distinguished in experiment with existing scanning force microscope techniques, and they should play significant roles in many other physical processes.
\end{abstract}

\maketitle

\pagebreak

The strain engineering provides an exciting approach to improve or manipulate various properties of graphene-based materials.\cite{NetoAHC} By generating strain, the experimentalists have been able to study the zero-field quantum hall effect in graphene and realize energy gap with the help of the stained  supperlattices,\cite{GuineaF2010} and observe the strain induced red-shift of the G mode.\cite{Mohiuddin,Huang} The strain effect has also been theoretically investigated for different aspects of graphene, such as the strain induced electrochemical potential and charges,\cite{KimEA,GuineaF2008} the electronic band gap for single layer graphene\cite{PereiraVM} and bilayer graphene\cite{ChoiSM}, the pseudomagnetic fields,\cite{GuineaF2010prb,LowT,KimKJ} the optical properties,\cite{PereiraVM2010} and the hydrogen storage capabilities of metal-decorated graphene.\cite{ZhouM} As a specific morphology of strain, a bubble in graphene was observed by Stolyarova {\it et.al} in 2009, which is formed due to the gas between graphene and its substrate.\cite{StolyarovaE} In 2010, Levy {\it et.al} found a very strong pseudo-magnetic field inside the bubble by using the scanning tunneling microscopy to do the spectroscopic measurements.\cite{LevyN} Quite recently, the Manchester's graphene research group show that it is possible to control the shape of the bubble in graphene by an external electric field.\cite{GeorgiouT} This technique is proposed to produce optical lenses with variable focal length.

In this paper, we investigate the effect of the bubble on the normal modes in graphene. We distinguish two opposite categories of normal modes, namely bright and dark modes, which are topologically different. The bubble is full of energy for the bright modes, while there is no energy inside the bubble in the dark mode. Due to their topological difference, the bright/dark mode is sensitive/insensitive to the shape of the bubble in the graphene. We examine the different behavior of bright and dark modes in the phonon thermal transport process. It is expected that the distinction between these two types of normal modes should be able to be measured in the experiment.

\begin{figure}[htpb]
  \begin{center}
    \scalebox{1.0}[1.0]{\includegraphics[width=8cm]{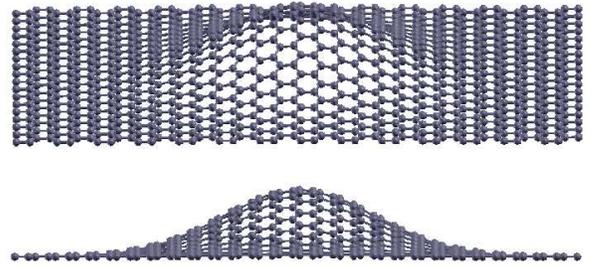}}
  \end{center}
  \caption{Two different views of a bubble in the graphene. The size parameters of the bubble are $(h, \lambda)=(7, 8)$~{\AA}, where $h$ is the height of the bubble and $\lambda$ reflects the lateral size of the bubble.}
  \label{fig_cfg}
\end{figure}
\begin{figure*}[htpb]
  \begin{center}
    \scalebox{1.0}[1.0]{\includegraphics[width=\textwidth]{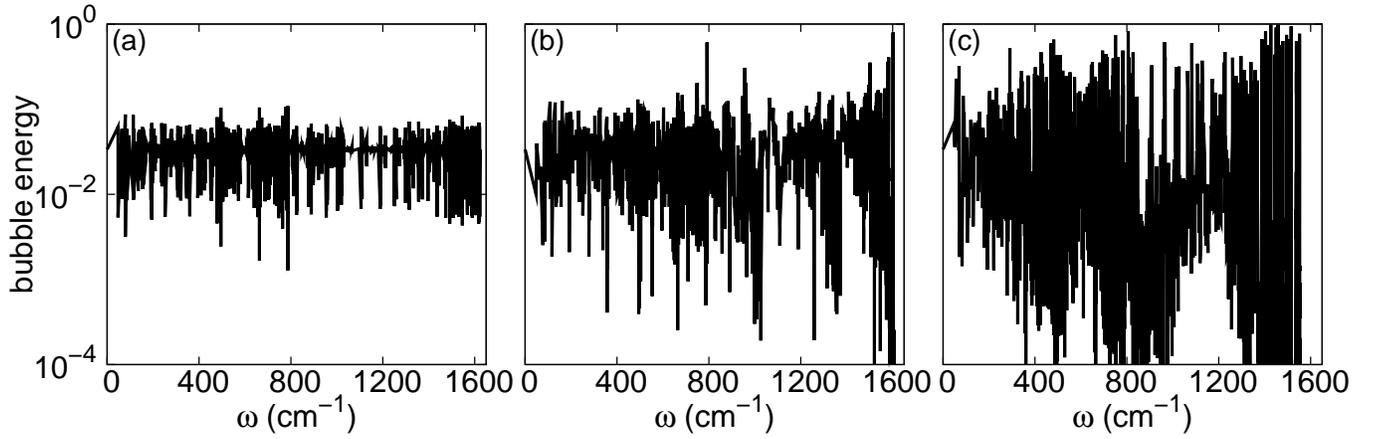}}
  \end{center}
  \caption{The energy within bubble regions for all normal modes in (a). pure graphene; (b). graphene with bubble (7, 8)~{\AA}; and (c). graphene with bubble (15, 8)~{\AA}. The $y$ axis is plotted in logscale.}
  \label{fig_eob}
\end{figure*}
\begin{figure*}[htpb]
  \begin{center}
    \scalebox{1.0}[1.0]{\includegraphics[width=\textwidth]{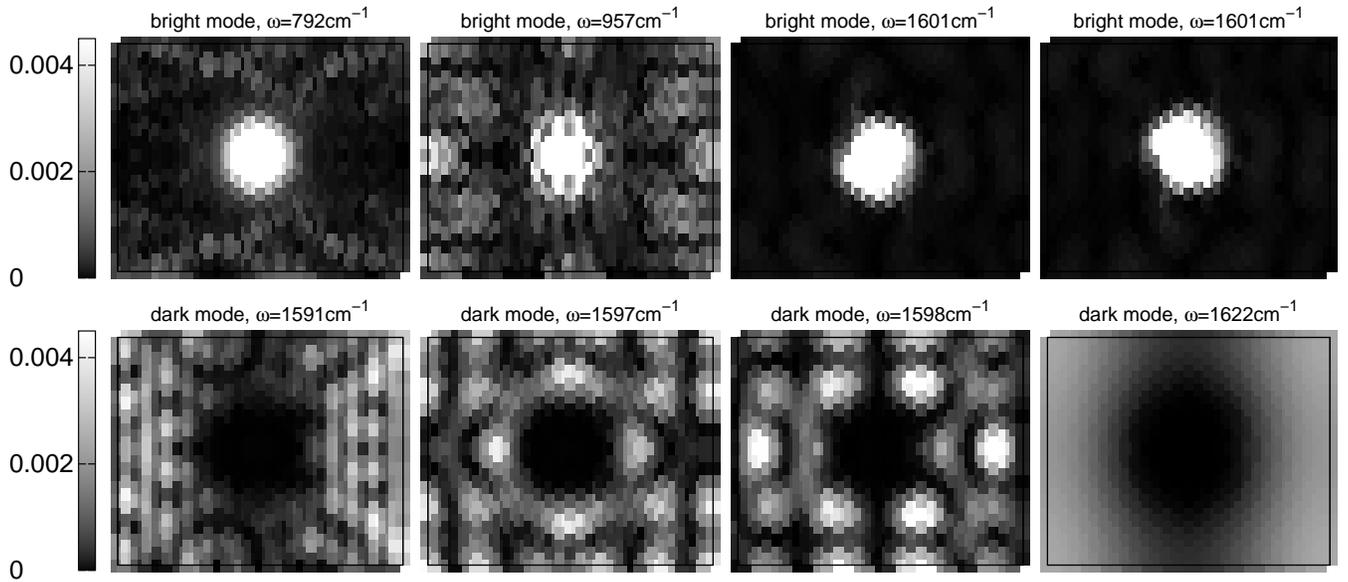}}
  \end{center}
  \caption{The energy spatial distribution of bright modes in the top four panels, and of dark modes in the bottom four panels for graphene with bubble (7, 8)~{\AA}. Value in each panel is the frequency of that normal mode.}
  \label{fig_u2}
\end{figure*}
Figure~\ref{fig_cfg} displays the configuration of a bubble within the graphene. The size of the graphene is 59.6~{\AA}$\times$44.3~{\AA}. The $xy$-plane lies in the graphene plane and $z$-direction is perpendicular to the graphene plane. The bubble has a Gaussian shape. The $z$-coordinate of atoms on the surface of the bubble is given by $z(x,y)=h*{\rm exp}\{-[(x-x_{0})^{2}+(y-y_{0})^{2}]/(2\lambda^{2})\}$, where $(x_{0}, y_{0})$ are the $x$- and $y$-coordinates of the center of the bubble. The bubble thus is characterized by the height $h$ and a lateral dimension $\lambda$. $(h,\lambda)=(7,8)$~{\AA} for the specific bubble shown in the figure.

We calculate the normal modes by diagonalizing dynamical matrix, which is obtained from the spring force constant model with longitudinal parameter $k_{l}$ and transverse parameter $k_{\perp}$ upto the fourth-nearest neighbors.\cite{SaitoR} The valence force field model\cite{JiangJW2008} has also been tried and the calculation results are qualitatively the same. Calculations shown below are based on the spring force constant model. For each normal mode $(\omega, \xi)$, the energy carried by atom $j$ is proportional to $\omega^{2}(\xi_{jx}^{2}+\xi_{jy}^{2}+\xi_{jz}^{2})$, where $\xi_{j\alpha}$ with $\alpha=x,y,z$ are the three components corresponding to atom $j$ in the polarization vector $\xi$. We analysis the energy spatial distribution for each normal mode. Fig.~\ref{fig_eob} shows the bubble energy in logscale for all normal modes in (a) pure graphene, (b) graphene with bubble (7, 8)~{\AA}, and (c) graphene with bubble (15, 8)~{\AA}. The bubble energy of a normal mode is defined to be the energy of this mode confined within a circular region of radius 5~{\AA} around the bubble center. The bubble energy has been normalized in such a way that the total energy of the normal mode is unitary. Panel (a) shows that the bubble energy of the pure graphene is about 3\% for all normal modes. It indicates that the energy is uniformly distributed on the entire graphene, considering the area of the bubble region is also around 3\% of whole area of the graphene. Panel (b) exhibits strong influence from the bubble on the energy spatial distribution of normal modes in graphene. On the one hand, the bubble energy of many normal modes is distinctly enhanced, and can reach as high as 80\% for some normal modes. On the other hand, the bubble energy of some other normal modes are strongly suppressed to be two or three orders smaller than the pure graphene's value of 3\%. These observations manifest that the bubble acts as an energy shield. It can thus split the normal modes in graphene into two opposite categories. One is to confine the energy inside the bubble region, while the other tries to repel energy for the bubble region. Panel (c) further reveals that the bubble energy is more seriously enhanced/suppressed by taller bubbles; thus they show a better performance as an energy shield.

\begin{figure}[htpb]
  \begin{center}
    \scalebox{1.05}[1.05]{\includegraphics[width=8cm]{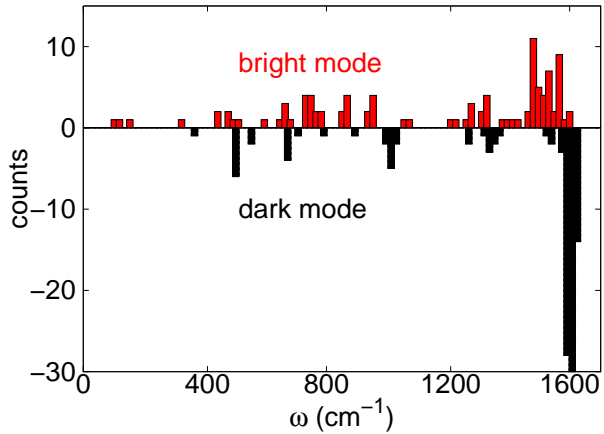}}
  \end{center}
  \caption{(Color online) The histogram of bright and dark modes in graphene with bubble (7, 8)~{\AA}. The total numbers of the bright and dark modes in this system are 101 and 112, respectively.}
  \label{fig_histogram}
\end{figure}
The energy spatial distributions are shown in Fig.~\ref{fig_u2} for eight normal modes in graphene with bubble (7, 8)~{\AA}. The four upper normal modes belong to the category with large bubble energy. The energy is concentrated so strongly in the bubble region that the bubble looks bright. For this reason, we name this type of normal modes as `bright modes'. The opposite behavior is observed in the four lower panels. The bubble energy is almost zero, which means that the bubble region is an energy vacancy in these modes. The bubble looks dark and similarly this type of normal modes are named `dark modes'. The current scanning force microscope technique is able to image the vibrational morphology of normal modes, such as the edge modes in the graphene nanoribbon.\cite{Sanchez} Hence, the bright and dark modes we have found here are readily to be observed in the experiment. The distinctness between these two type of normal modes is also quite possible to be verified experimentally. Fig.~\ref{fig_histogram} locates the position of bright and dark modes in the frequency space for graphene with bubble (7, 8)~{\AA}. It shows that the bright mode is more uniformly distributed in whole frequency range while dark modes mainly locates in the high frequency region. This figure indicates that bright mode will play a role in whole temperature range. However, the dark mode will become important at high temperatures, where high frequency normal modes can be sufficiently excited.

\begin{figure}[htpb]
  \begin{center}
    \scalebox{0.9}[0.9]{\includegraphics[width=8cm]{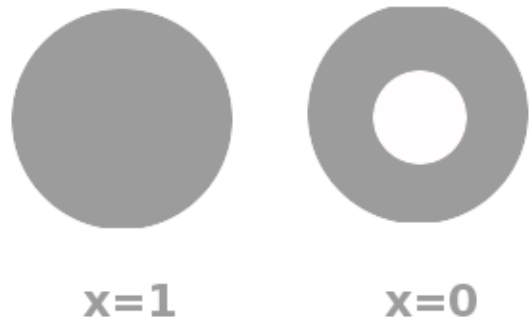}}
  \end{center}
  \caption{The circular area (left) with the Euler characteristic of 1 and the cirque area (right) with the Euler characteristic of 0.}
  \label{fig_topology}
\end{figure}
From the topological point of view, we can clearly distinguish the bright and dark modes. The topological nature of the bright and dark modes is equivalent to the circular and cirque respectively as shown in Fig.~\ref{fig_topology}. The gray filled area corresponds to the high density of spatial energy for normal modes. The Euler characteristic is the topological invariant. It coincides with the Euler number for the two geometries in Fig.~\ref{fig_topology}. The Euler characteristic ($\chi$) can be calculated through the Gauss-Bonnet theorem, after applying to these two special geometries:
\begin{eqnarray}
\chi=\frac{1}{2\pi}\int_{edge}K_{g}ds,
\end{eqnarray}
where $K_{g}$ is the curvature and $ds$ is the line element along the edge. Using this formula, we get the Euler characteristic for the circular and the cirque are 1 and 0, respectively. Hence, the Euler characteristic for the bright mode is 1 and that of the dark mode is 0.

From the above, we have learn the energy spatial distribution for the bright and dark modes, from which we demonstrate that these two types of normal modes are of different topological nature with different Euler characteristic. As we know, the topological invariant is of key importance in many physical processes. It is quite possible that the different topological invariant of the bright and dark modes will lead to observable distinctness in those physical processes where the normal mode is an important exciton. The phonon thermal transport is one of such physical process,
\begin{figure}[htpb]
  \begin{center}
    \scalebox{1.0}[1.0]{\includegraphics[width=8cm]{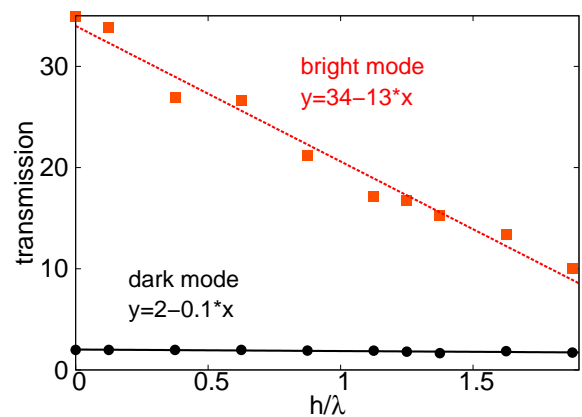}}
  \end{center}
  \caption{(Color online) The transmission for bright and dark modes versus the shape of the bubble. The two straight lines are the fitting curves for two data sets.}
  \label{fig_size_transmission}
\end{figure}
\begin{figure}[htpb]
  \begin{center}
    \scalebox{1.0}[1.0]{\includegraphics[width=8cm]{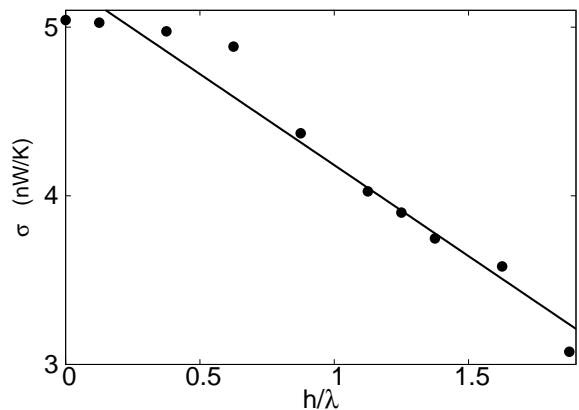}}
  \end{center}
  \caption{The thermal conductance of graphene with bubbles of varying shape at room temperature. The straight line is guide to the eye.}
  \label{fig_size}
\end{figure}
 where normal mode is the energy carrier. The thermal transport in graphene is currently an active field, where very high thermal conductivity was found.\cite{Balandin,NikaPRB,NikaAPL} For a recent review of this subject, we refer to Ref.~\onlinecite{Balandin2011}. We apply NEGF approach\cite{WangJS} to calculate the transmission function, $T[\omega]$, of normal modes.
\begin{eqnarray}
T[\omega] & = & {\rm Tr}\left(G^{r}\Gamma_{L}G^{a}\Gamma_{R}\right),
\end{eqnarray}
where $G^{a}=\left(G^{r}\right)^{\dagger}$ is the advanced Green's function and $\Gamma_{L/R}$ are the two self energy. The phonon thermal conductance is calculated by the Landauer formula. Fig.~\ref{fig_size_transmission} shows the transmission for a bright mode at $\omega=792$cm$^{-1}$ and a dark mode at $\omega=1622$cm$^{-1}$ in graphene with bubbles of different shapes. Obviously, the transmission for the bright mode is very sensitive to the geometrical ratio $h/\lambda$, and decreases linearly with the increase of $h/\lambda$. It indicates that this mode becomes more bright for a bubble of lager $h/\lambda$, where the energy is more concentrated in the bubble region. On the contrary, we find that the dark mode is insensitive to the ratio of $h/\lambda$. This is because the energy of the dark mode is distributed throughout the whole graphene sample except the bubble region. Thus it is insensitive to the local environment of the bubble. The room-temperature thermal conductance result in Fig.~\ref{fig_size} confirms that the bright mode plays important role and leads to a linearly decreasing of the thermal conductance with the increase of $h/\lambda$.

In conclusion, we have studied the normal modes in graphene with bubbles and found two topologically different types of normal modes, namely the bright and dark modes. The bright modes confine energy inside the bubble region and has an Euler characteristic of 1; while the dark mode repel energy for bubble region and has an Euler characteristic of 0. As a result of their different topological nature, the bright mode is sensitive to the shape of the bubble; yet the dark mode is insensitive to the shape of the bubble. The distinctness between bright and dark modes is expected to be observed in the experiment. We have also examined the difference between these two types of normal modes in the phonon thermal transport process, and find that the transmission of bright/dark mode is sensitive/insensitive to the geometrical ratio $h/\lambda$.

\textbf{Acknowledgements} The authors thank X. F. Xu and L. F. Zhang for helpful discussion. The work is supported by a URC grant of R-144-000-257-112 of National University of Singapore.

\end{document}